\documentclass[11pt,titlepage]{article}
\usepackage{graphics,epsfig,color,latexsym}
\usepackage{graphicx}
\usepackage{dcolumn}
\usepackage{bm}
\usepackage{epsfig}
\usepackage{wrapfig}

\newcommand{\bi}{\bibitem}

\newcommand{\Sslash}{\kern 0.2 em S\kern -0.56em \raisebox{0.3ex}{/}}

\newcommand{\bgi}{\begin{itemize}}
\newcommand{\eni}{\end{itemize}}

\newcommand{\bea}{\begin{eqnarray}}
\newcommand{\eea}{\end{eqnarray}}

\newcommand{\bb}{\begin{thebibliography}}
\newcommand{\eb}{\end{thebibliography}}

\newcommand{\be}{\begin{eqnarray}}
\newcommand{\ba}{\begin{array}}
\newcommand{\ea}{\end{array}}
\newcommand{\ee}{\end{eqnarray}}

\newfont{\fib}{cmfi10 at 10pt}

\newcommand{\eg}{{\it e.g.}\ }

\begin{document}
\begin{titlepage}
\begin{center}
\begin{title} 
\bf{\Large{Transversity and Transverse Spin in Nucleon Structure through 
 SIDIS \\
at Jefferson Lab
}}
\\
\vspace{2cm}
{\bf  Submitted for \linebreak 
Nuclear Physics Long Range Plan ``QCD and Hadron Physics''} \\
Date: 2/13/2007\\

\vspace{2cm}
A. Afanasev$^{1}$, M. Anselmino$^{2}$, H. Avakian$^{3}$, G. Cates$^{4}$, 
J.-P. Chen$^{3}$, E. Chudakov$^{3}$, E. Cisbani$^{5}$,\\
C. de Jager$^{3}$, 
L. Gamberg$^{6}$, H. Gao$^{7}$, F. Garibaldi$^{5}$, X. Jiang$^{8}$, 
K. S. Kumar$^{9}$, Z.-E. Meziani$^{10}$, \\
P. J. Mulders$^{11}$, J.-C. Peng$^{12}$, 
X. Qian$^{7}$, M. Schlegel$^{3}$, P. Souder$^{13}$, 
F. Yuan$^{14}$, L. Zhu$^{12}$ \\
\vspace{0.3cm}
\it
$^{1}$Hampton University, Hampton, VA 23668, USA \\
$^{2}$Universita di Torino and INFN, Sezione di Torino, I-10125, Torino, Italy \\
$^{3}$Jefferson Lab, Newport News, VA 23606, USA \\
$^{4}$University of Virginia, Charlottesville, VA 22901, USA \\
$^{5}$INFN, Sezione di Roma III, 00146 Roma, Italy \\
$^{6}$Penn State-Berks, Reading, PA 19610, USA \\
$^{7}$Duke University,Durham, NC 27708, USA \\ 
$^{8}$Rutgers University, Piscataway, NJ 08855, USA \\
$^{9}$University of Massachusetts, Amherst, MA 01003, USA \\
$^{10}$Temple University, Philidalphia, PA 19122, USA \\
$^{11}$Vrije Universiteit, NL-1081 HV Amsterdam, the Netherlands \\
$^{12}$University of Illinois at Urbana-Champaign, Urbana, IL 61801, USA \\
$^{13}$Syracuse University, Syracuse, NY 13244, USA \\
$^{14}$Brookhaven National Laboratory, Upton, NY 11973, USA \\
 
\end{title}
\end{center}
\end{titlepage}

\subsection{Introduction}

Spin is an intrinsic quantum-mechanical and relativistic property of 
the constituents of matter which plays a fundamental role in physical
processes and in theories of fundamental interactions. Three decades
of intensive experimental and theoretical investigation has resulted in 
a great deal of knowledge on the partonic origin of the nucleon spin structure.
In particular, considerable knowledge has been gained 
from inclusive deep-inelastic scattering (DIS) experiments 
at  CERN, SLAC, DESY and more recently at JLab and 
RHIC on the longitudinal structure -- the $x$-dependence  and 
the helicity distributions -- in terms of the 
unpolarized (denoted $q(x)$ or $f_1^q(x)$) and helicity (denoted $\Delta q(x)$
or $g_1^q(x)$) 
parton distribution functions for the various flavors (indicated by $q$).
We have also learned that precise knowledge of the transverse structure
is an essential part of the partonic spin and momentum substructure 
of the nucleon. 
This information can only be obtained by a consideration
of data beyond those obtained from the unpolarized and longitudinally polarized processes.
Theoretically, this entails the exploration of the QCD parton model
 beyond the collinear approximation.

Indeed, the 
extension of the QCD parton model beyond the collinear approximation 
and the systematic study of semi-inclusive deep-inelastic lepton nucleon
scattering (SIDIS) has emerged as an essential tool to probe both 
the nucleon's longitudinal and transverse momentum and spin structure. 
The azimuthal dependence in the scattering of leptons 
off transversely polarized nucleons 
is explored through the analysis of transverse single spin 
asymmetries (TSSAs). QCD predicts that these observables are 
factorized convolutions of leading-twist 
transverse momentum dependent parton distributions (TMDs) and 
fragmentation functions~\cite{mulderstang,boermul,jima,col_metz}. 
These functions provide {\em essential non-perturbative} 
information on the partonic sub-structure of the nucleon; they 
offer a rich understanding of the motion of partons inside the 
nucleon, of the quark orbital properties and of spin-orbit correlations.

The Thomas Jefferson National Accelerator Facility (JLab)
is exceptionally suited to carry out precision studies 
of the longitudinal and
transverse spin and momentum structure of the  nucleon 
as well as its flavor
decomposition from SIDIS experiments due to 
its high luminosity in combination with large acceptance detectors and
the kinematic coverage in longitudinal and transverse momentum.

\subsection{Transversity and TMDs through SIDIS}

In recent years new theoretical~\cite{jaffeji,rat} 
and experimental efforts~\cite{hermes,compass} have been devoted to explore 
the transverse 
spin and momentum structure of the nucleon. 
This concerns in particular the investigations of the chiral-odd 
transversely polarized quark distribution function (or transversity, denoted as $\delta q(x), h_1^q(x)$ or 
also $\Delta_T q(x)$~\cite{artru}). Like the axial charge $\Delta q =\int_0^1 dx\ (g_1^q(x) +\bar g_1^q(x))$,
the nucleon tensor charge $\delta q =\int_0^1 dx (h_1^q(x)-\bar{h}^q_1(x))$
is a basic property of the nucleon, which is dominated by valence quarks.
It is crucial 
to note that due to its
chiral-odd property, $h_1^q$ does not mix with gluons under evolution
and it receives no contribution from quark-antiquark pairs in the 
{\em sea}. Thus,  it
is a non-singlet {\em valence-dominated} distribution function. 
 Transversity,
 together with the unpolarized and helicity parton distribution functions,
 is one of the three basic functions describing the distribution
 of quarks in the nucleon. 
It satisfies the Soffer bound~\cite{soffer}, 
$\left|2h_1^q( x,{\scriptstyle Q^2})\right|\le f_1^q(x,{\scriptstyle Q^2}) 
+ g_1^q(x,{\scriptstyle Q^2})$
for each 
flavor at the scale ${\scriptstyle Q^2}$. Its meaning is transparent
in a transverse spin basis: $h_1$ describes
 the probability to find a quark
with spin polarized along the transverse spin of a polarized
nucleon minus the probability to find it polarized 
oppositely~\cite{jaffeji}. In terms of a helicity description,
the transversity distribution is chiral odd, 
that is, the helicities of both the quark and nucleon  flip. 
Thus, due to the helicity
conservating property of the QCD interactions 
$h_1$ decouples at leading twist in an expansion of inverse powers
of the hard scale in inclusive deep-inelastic scattering. 
However, paired with
another hadron in the initial state,  \eg Drell-Yan 
Scattering~\cite{ralston}, 
or in the final state, \eg semi-inclusive deep-inelastic~\cite{colnpb} 
scattering, the transverse structure of the nucleon 
can be accessed without suppression by a hard scale. 
Thus, at leading twist $h_1$ can be accessed
with another distribution or fragmentation function.

The most feasible  way to access the transversity distribution
 function  is via an azimuthal single spin asymmetry, 
 in semi-inclusive deep-inelastic
lepto-production of mesons on a transversely polarized nucleon target, 
$e\, N^\uparrow\rightarrow e\, \pi \, X$.  In this case 
the other chiral-odd partner is the Collins~\cite{colnpb} fragmentation
function, $H_1^\perp$. 
Schematically from  factorized QCD,  this transverse single
spin asymmetry (TSSA) contains $h_1$ and $H_1^\perp$, 
$A_{UT}\sim h_1\otimes H_1^\perp$
($U\equiv$ unpolarized electron
beam, $T\equiv$ transversely polarized target)~\cite{boermul}.

The first evidence 
of non-zero transversity has
been observed in the HERMES experiment~\cite{hermes}
where an unpolarized electron beam is scattered off
a transversely polarized proton in SIDIS,
$e\, p^\uparrow\rightarrow e^\prime\, \pi \, X$,
whereas the COMPASS measurements~\cite{compass}
of the Collins and Sivers asymmetries on the deuteron are 
consistent with zero within experimental uncertainties.
By contrast to inclusive deep-inelastic lepton-nucleon scattering 
where transverse momentum is integrated out,
these processes are sensitive to 
 transverse momentum scales on the order of the intrinsic quark momentum
$P_T\sim k_\perp$.  
This is evident by considering the generic structure of the TSSA  
for a transversely polarized nucleon target  
which is characterized by interference between helicity flip and helicity
non-flip amplitudes  $ A_{UT}\sim {\rm Im}( f^{*\, +}f^-)$.  In 
the collinear limit of QCD partonic processes
conserve helicity and Born amplitudes are real~\cite{kane}. Thus, 
a reaction mechanism to account for  the interference
between helicity flip amplitudes at leading twist requires a
 probe of the nucleon at a scale sensitive  to 
the intrinsic quark transverse momentum. This 
is roughly set by the confinement scale  
$k_\perp\sim\Lambda_{\rm qcd}$~\cite{ansel}.

Transverse spin  asymmetries arise from
{\em naive time reversal odd} ($T$-odd) correlations
between transverse spin $\bm{S}_T$, longitudinal momentum $\bm{P}$ 
and intrinsic quark momentum $\bm{k}_\perp$~\cite{sivers,colnpb}. 
This is  depicted by the generic vector product 
$i\bm{S}_T\cdot\left({\bm P}\times {\bm k}_\perp\right)$.
These correlations 
imply  that the reaction mechanism is associated with  
leading twist-two helicity flip, $T$-odd transverse momentum dependent (TMD)
parton distribution~\cite{sivers} and fragmentation~\cite{colnpb} functions
(PDFs \& FFs). These ingredients enter the 
factorized~\cite{jima,col_metz} hadronic tensor
for semi-inclusive deep-inelastic scattering.

A crucial theoretical breakthrough~\cite{colplb,bel,bpm}  
was that the reaction 
mechanism is due to non-trivial phases arising  from 
the color gauge invariant property of QCD. 
This leads to the picture 
that TSSAs arise from   initial and 
final state interactions~\cite{bhs,jiyuan,ggoprh} (ISI/FSI)
 of the active quark with the  soft distribution or fragmentation
remnant in SIDIS. 
Thus, $T$-odd TMDs are of crucial importance because they
 posses transversity properties
as well as the necessary phases to
account for   TSSAs at leading twist.

Exploring the transverse spin structure 
of the TMD PDFs  reveals evidence of a rich
spin-orbit structure of the nucleon.
When the transverse spin-momentum correlations are 
 associated with the nucleon where the quark remains {\em unpolarized}, 
the so-called Sivers function associated 
with the helicity flip of a transverse spin  nucleon target 
dominates at leading twist. 
Since the quark is unpolarized
in the Sivers function, 
the orbital angular momentum
of the quarks must come into play to conserve overall angular momentum
in the process~\cite{burk}. 
This result has a far-reaching impact on the QCD theory of generalized angular
momentum~\cite{brodgard}.

\begin{figure}[t]
\begin{center}
\includegraphics[width=0.55\textwidth,bb= 10 140 540 660,angle=-90]
{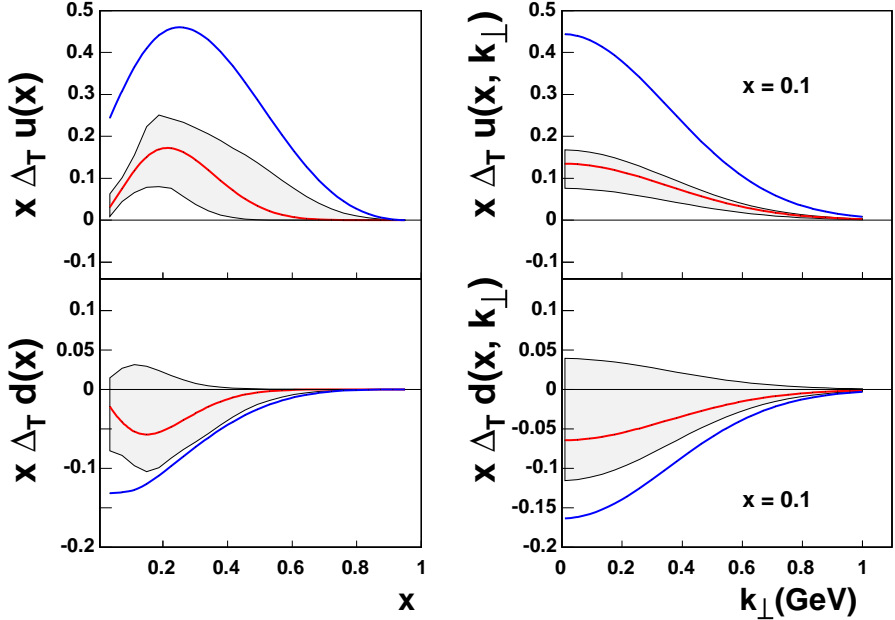}
\caption{\label{fig:transv}
The transversity distribution functions $\Delta_T q(x)$ (also denoted as 
$\delta q(x)$ or $h_1^q(x)$)
for $u$ and $d$ quarks as determined
from the global best fit analysis~\cite{anseltrans}.
Left panel, $x\,\Delta _T u(x)$ 
and $x\,\Delta _T d(x)$ upper and lower plot,  shown as
functions of $x$. The Soffer bound \cite{soffer},  
bold blue line. Right panel, the unintegrated transversity
distributions, $x\,\Delta _T u(x,k_\perp)$ and
$x\,\Delta _T d(x,k_\perp)$ upper and lower plot, 
as defined in~\cite{anseltrans}, as
functions of $k_\perp$ for fixed $x$. 
}
\end{center}
\end{figure}

When the single transverse spin-momentum correlation
 is associated with the fragmenting quark,
we access  the Collins fragmentation 
function.  Indeed as stated earlier  the Collins function
couples to the chiral-odd transversity distribution function $h_1(x)$. 
Understanding its properties and accessing it
through experiment is a crucial ingredient to pin down $h_1$.

First attemps~\cite{anseltrans} have been performed to extract 
the 
transversity distributions (up to a sign) 
by combining SIDIS~\cite{hermes,compass} data with
 $\rm e^+ e^-$ data~\cite{belle} on the Collins function 
(see Fig.~\ref{fig:transv}). Within the large errors, 
the Soffer bound
is respected.

Complementing the 
new data from the HERMES~\cite{hermes}, COMPASS~\cite{compass},
 and BELLE~\cite{belle} experiments, 
 the approved Hall A experiments E06-010/E06-011~\cite{jlabhalla} on the 
neutron (with polarized $^3$He)
will facilitate a flavor decomposition 
 of the transversity distribution 
function, $h_1$~\cite{jaffeji,rat}
and the first moment of
the $T$-odd Sivers distribution function 
$f_{1T}^{\perp (1)}$~\cite{sivers} 
in the proposed but limited kinematic 
regime with limited precision.
However a model-independent determination of these leading twist functions 
demands a wider kinematic range with high precision in {\em three dimensions} 
($x, z, \bm{P}_T$).

\subsection{Three Dimensional Kinematic Map of TMDs:
JLAB 12 GeV with Large Acceptance}

The prospect of mapping out the nucleon's partonic substructure through
TMDs from SIDIS experiments is encouraging.
The endeavor to determine the nucleon's spin-momentum structure from SSAs 
in SIDIS demands high-precision data in longitudinal and transverse 
momentum kinematics ($x$, $z$, and $\bm{P}_T$).

The JLab $12\ \rm GeV$ 
upgrade with the CLAS12 detector (with polarized proton and deuteron targets), 
and the proposed 
solenoid large-acceptance detector 
(with a polarized neutron target) in Hall A 
provide an unprecedented opportunity to obtain  a 
{\em three-dimensional} map of the Collins and
Sivers asymmetries in the kinematical region
$0.1\le x \le 0.5$, $0.3 \le z \le 0.7$ with $P_T \le 1.5\ {\rm GeV}$,
 necessary to precisely 
determine the nucleon's partonic substructure.

\begin{figure}[htb]
\centering
\epsfig{file=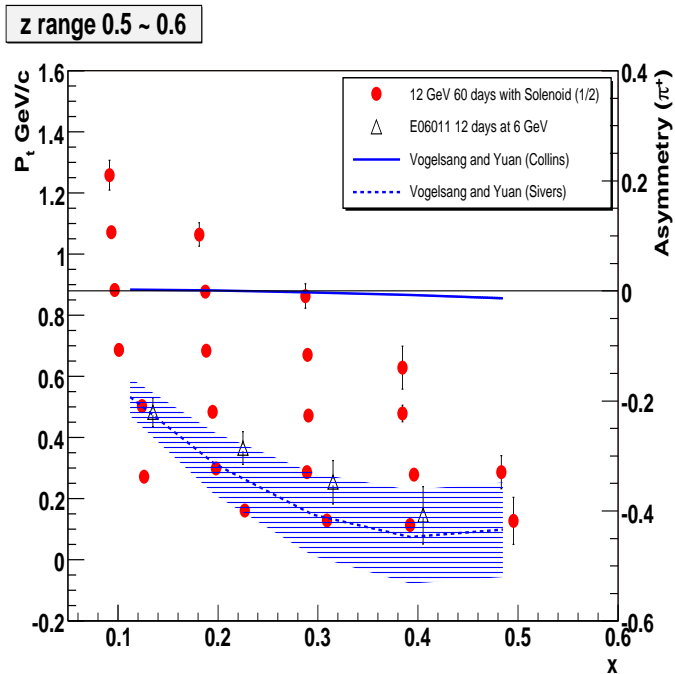,height=3.6in,width=3.6in}
\caption{The projection of the Sivers and the Collins asymmetry measurement 
for $\pi^+$ production on the neutron with
the proposed selonoid detector for $z =0.5 - 0.6$ 
  as a function of $P_T$ and $x$.
 The size of the error bar follows the axis on the right
  while the left axis shows the $P_T$ range. The solid (dashed) line
  shows a model prediction~\cite{yuan} of the Collins (Sivers) asymmetry.}
\label{fig:pip}
\end{figure}

Fig.~\ref{fig:pip} shows a sample plot of the projection of
the Sivers and the Collins asymmetry measurement 
for $z =0.5 - 0.6$ as a function of $P_T$ and
  $x$
for 60 days of running using the solenoid detector with 
a transversely polarized $^3$He (neutron)
target detecting $\pi^+$ in coincidence with the
scattered electrons. 
Also shown in the figure are theoretical predictions from a model~\cite{yuan} 
which fits the HERMES data.

This new precision measurement on the neutron together with the CLAS12 measurement on the proton and the deuteron and the world data
will allow for the first time an extraction of the Sivers 
distribution function, the transversity distribution function as well as 
the Collins fragmentation function. The Collins function can be cross-checked with the $e^+e^-$
annihilation measurement. 
This will provide an accurate determination 
of the transversity distribution function $h_1$, its first $x$ moment -- the 
nucleon tensor charge, $\delta q$ --  
as well as the flavor dependence of the first 
$\bm{k}_\perp$-moment of the Sivers function, $f_{1T}^{\perp (1)}$. The tensor
charge is a fundamental property of the nucleon, similar to electric charge
or total spin, and can be calculated from Lattice QCD.

\subsection{ Long Range Impact}
As stated in the pre-conceptual design report (pCDR) for 
the $12 \rm GeV$ upgrade
of JLab~\cite{jlabpcdr} the measurement of TSSAs in SIDIS 
is a central tool to measure the leading twist transverse momentum dependent
distribution and fragmentation (TMD) functions.
In particular, the Collins mechanism provides access to 
 the {\em valence-dominated} 
quark transversity function and the nucleon's tensor charge,
 while the Sivers mechanism provides
information on the spin-orbit interactions of quarks within the nucleon.

The theoretical and experimental endeavor to analyze the
Sivers and Collins asymmetries serves the purpose
of exploring quantum chromodynamics (QCD)
beyond the collinear limit {\em down} to  transverse momentum 
scales $P_T\sim \Lambda_{\rm qcd}$.  This provides an 
unprecedented three-dimensional kinematic map of the spin-momentum
partonic sub-structure of the nucleon.
Recent evidence of the Collins and Sivers effect from the HERMES~\cite{hermes},
COMPASS~\cite{compass} and BELLE~\cite{belle} 
experiments have set the groundwork for 
the next phase at performing high-precision measurements.

The JLab 12 GeV upgrade with the proposed solenoid
detector and the CLAS12 detector
will provide the granularity and three-dimensional 
kinematic coverage in longitudinal and transverse momentum,
$0.1\le x \le 0.5$, $0.3 \le z \le 0.7$ with $P_T \le 1.5\ {\rm GeV}$
to measure  these leading twist 
chiral-odd and  $T$-odd  
quark distribution and fragmentation functions. 
We close by emphasizing that  due to the large
$x$ experimental reach of these detectors JLab 12 GeV 
is the {\em ideal setting}
to obtain precise data on the {\em valence-dominated} transversity 
distribution function and the tensor charge.

\vskip 0.2 cm

\section*{Acknowledgments}
This work is supported in part by 
the U.S. Department of Energy under contracts, DE-AC05-84ER40150,
modification No. M175, under which the Southeastern Universities
Research Association operates the Thomas Jefferson National Accelerator
Facility, DE-FG02-07ER41460 (L.G.), and  DE-FG02-03ER41231 (H.G.).

\end{document}